\documentclass[
 reprint, 
 amsmath,amssymb, aps, prb, 
 ]{revtex4-2}
\usepackage{placeins}
\usepackage{graphicx}
\usepackage{dcolumn}
\usepackage{bm}
\usepackage{float}
\begin{document}

\preprint{APS/123-QED}

\title{Superheating Field of Extreme Type-II Superconductors Calculated from the Eliashberg Theory}
\author{Ramesh Paudel$^{1,\ast}$}
\author{Alex Gurevich$^{1,\ast}$}
\affiliation{$^1$Department of Physics, Old Dominion University, Norfolk, Virginia 23529, USA}
\thanks{Contact author: rpaudel@odu.edu; gurevich@odu.edu}%
\date{\today}

\begin{abstract}

We calculate the superheating field $H_s$ of a type-II superconductor with a large Ginzburg-Landau parameter from the  
isotropic Eliashberg theory. It is shown that the temperature dependencies of $H_s(T)$ can be obtained from a thermodynamic approach for different  
scattering parameters $p=\hbar/2\pi\tau_Nk_BT_c$ in a wide range of the nonmagnetic impurity scattering times $\tau_N$ 
ranging from the clean $(p\ll 1)$ to the dirty $(p\gg 1)$ limits. Specific calculations were done for Nb and Nb$_3$Sn using their 
electron-phonon spectral functions extracted from tunneling measurements. We show that the ratio $H_s(T,p)/H_c(T)$ calculated from the Eliashberg theory exceeds that of the weak coupling BCS model, where $H_c$ is the thermodynamic magnetic field.  We also show that, even though nonmagnetic impurities are pairbreakers in the current-carrying state, $H_s(0,p)$ for Nb$_3$Sn at $T=0$ has a maximum at $p\approx 0.2$ at which the ratio $H_s/H_c$ is about $6\%$ higher than that in the BCS clean limit.  
\end{abstract}

\maketitle

\section{\label{sec:level1}Introduction}

Stable vortex-free Meissner state in type-II superconductors exists at the applied magnetic field $H$ smaller than the lower critical magnetic field $H_{c1}$. Yet the Meissner state can remain metastable at much higher magnetic fields $H_{c1}<H<H_s$, up to a superheating field $H_s$ at which the Bean-Livingston surface barrier \cite{Bean} for penetration of vortices disappears. The superheating field is a fundamental field limit for a weakly dissipative state under strong  low-frequency electromagnetic field $\hbar\omega\ll \Delta$ at $T\ll T_c$, where $\Delta$ is the superconducting gap and $T_c$ is the critical temperature. There is a renewed interest in the physics of $H_s$ due to advances in superconducting resonators for particle accelerators \cite{accel1,accel2} and free electron lasers~\cite{Xlasers} which require extremely high quality factors $Q\sim 10^{10}-10^{11}$ at strong radio-frequency (RF) fields. For instance, the state-of-the-art Nb cavities can operate in the Meissner state at 2K and the surface RF fields exceeding the thermodynamic critical field $B_c=\mu_0H_c\approx 200$ mT at which the screening current density approaches the depairing limit \cite{accel1}. 

The dc superheating field $H_s$ at $T\approx T_c$ was calculated from the Ginzburg-Landau (GL) theory by many authors \cite{sh1,sh2,sh3,sh4}. It has been shown that, as the applied field $H$ reaches $H_s$, the screening current density at the surface $J(x)$ becomes of the order of the depairing current density $J_d\simeq  H_c/\lambda_L$ and the Meissner state gets unstable with respect to small perturbations of current and the order parameter which have the wavelength $\sim (\xi^3\lambda_L)^{1/4}$ along the surface and decay over the length $\sim (\lambda_L\xi)^{1/2}$ perpendicular to the surface, where $\lambda_L$ is the London penetration depth and $\xi$ is the coherence length. Such perturbations describe the initial stage of penetration of a periodic vortex row.  The superheating instability of inhomogeneous screening current density results in $H_s$ decreasing with the GL parameter $\kappa=\lambda_L/\xi$, from $H_s\approx 1.2H_c$ at $\kappa=1$ to $H_s=0.745H_c$ at $\kappa\gg 1$. In the latter case $H_s$ can be obtained from the condition that $J(x)$ at the surface reaches $J_d$ for a uniform current flow. 

Calculation of $H_s(T)$ at low temperatures requires solving the nonlinear Eilenberger equations \cite{eilenb} taking into account pairbreaking effects which can be rather different from those at $T\approx T_c$  \cite{Bardeen, Parmenter1,Parmenter2,Maki1,Maki2,Fulde}. The first calculation of $H_s(T)$ for the entire temperature range $0<T<T_c$ in the clean limit and $\kappa\to \infty$ was done by Galaiko \cite{Galaiko} who obtained $H_s=0.84H_c$ at $T\rightarrow 0$, and $H_s= 0.745H_c$ at $T\rightarrow T_c$. Subsequent calculations of $H_s(T)$ at $0<T<T_c$ in the clean limit at $\kappa\to\infty$ showed that $H_s(T)$ has a maximum at low $T$ \cite{Catelani}. The effect of magnetic and nonmagnetic impurities on $H_s(T)$ at $\kappa\gg 1$ was calculated in Ref. \cite{LG} in which a maximum in $H_s$ as a function of concentration of impurities was found.  Calculation of $H_s$ outside the GL region at finite $\kappa$ requires taking into account complex interplay of nonlinear pairbreaking and nonlocality of the electromagnetic response. Recently $H_s$ was calculated from the Eilenberger equations for a clean superconductor with $\kappa\sim 1$ \cite{kub}. All microscopic calculations of $H_s$ have shown that $H_s$ at $T\ll T_c$ exceeds $H_s$ extrapolated from the GL region. 

Given the importance of $H_s$,  several ways of increasing $H_s$ by surface nanostructuring have been proposed, including deposition of high-$T_c$ superconducting multilayers  \cite{ml1,ml2,ml3,ml4,ml5} or a dirty overlayer with a higher impurity concentration at the surface \cite{sauls}. The superheating fields of such structures have been evaluated using the Eilenberger and Usadel equations in the limit of $\kappa=\lambda_L/\xi\gg 1$ \cite{ml1,ml2,ml3,ml4,ml5,sauls} and from the GL theory at a finite $\kappa$ \cite{PG}. In all these cases $H_s$ can exceed the respective superheating fields of multilayer components at an optimal layer thickness maximizing the current counterflow induced by the superconducting substrate in the top layer. These effects can be used to optimize high-field RF losses in superconducting resonators \cite{ag}. 

Evaluation of $H_s$ in real material is complicated not only by the complexity of the stability analysis of Meissner state in the Eilenberger theory but also by the fact that it is based on the BCS model which cannot reliably predict $H_s$ in such superconductors as Pb, Nb or A15 materials with strong electron-phonon interaction. Such superconductors are described by the Eliashberg theory \cite{C,MC} which has been advanced in recent years by the  incorporation of ab initio DFT calculations taking into account realistic electronic band structures, the effects of crystal anisotropy and the Coulomb electron interaction \cite{ab1,ab2}.  This approach has been used to describe superconductivity in superhydrides \cite{hydr1,hydr2} or to search for room-temperature superconductors \cite{rts}. Recently it was also used to predict superconductors with high critical fields $H_{c2}$ and $H_{c1}$ using machine-learning algorithms \cite{highH} but the superheating field has not yet been calculated.

The Eliashberg theory brings about essential features missing in the Eilenberger quasiclassical theory such as electron-phonon strong-coupling corrections and the renormalization of the electronic band parameters which are expected to increase $H_s$ as compared to the BCS predictions. At the same time, inelastic electron-phonon collisions result in finite quasiparticle lifetimes and pairbreaking subgap quasiparticle states which reduce $H_s$. Addressing the interplay of these effects in $H_s$ is the goal of this work in which we calculate $H_s$ from the conventional isotropic Eliashberg theory in the limit of $\kappa\gg 1$. Given that Nb and Nb$_3$Sn are currently the best-performing superconductors used in high-field resonators~\cite{accel1,Posen,Keckert, Eremeev}, we calculated $H_s$ for Nb$_3$Sn and a dirty Nb using their electron-phonon spectral functions extracted from tunneling measurements. It is shown that the so-calculated $H_s$ exceeds $H_s$ obtained from the Eilenberger theory. We also calculated the effect of nonmagnetic impurities on $H_s$ which turns out to be rather nontrivial. Indeed, while nonmagnetic impurities do not affect $T_c$ and $H_c$ at zero superfluid velocity, they become pairbreakers in the current-carrying state \cite{Maki1,Maki2}, which results in a nonlinear Meissner effect and intermodulation~\cite{nlme}. Yet we show that a small concentration of nonmagnetic impurities can increase $H_s$ as compared to $H_s$ of a clean superconductor. 

The paper is organized as follows: In Sec. II, we present the Eliashberg equations which are used to calculate the superfluid current density as a function of the superfluid velocity. Sec. III contains results of our numerical calculations of the superheating field, thermodynamic critical field, depairing current density and the London penetration as functions of temperature and nonmagnetic scattering time for Nb and Nb$_3$Sn. We summarize our conclusions in Sec. IV. 

\section{Eliashberg Equations}

In this paper we calculate $H_s$ at $\kappa\gg 1$ for which the screening current density $J(x)$ decreases slowly over $\xi$. In this case $H_s$ is determined by a local dependence of $J(A(x))$ on the vector potential $A(x)$ which is calculated by solving  the isotropic Eliashberg equations in the presence of a uniform superflow \cite{NC}: 
\begin{gather}
\tilde{\Delta}_n
= \pi T \sum_{m=-\infty}^{\infty}
\left[\lambda(n-m)-\mu^\ast\theta(\omega_c-|\omega_m|)\right] \nonumber \\
\times \int_{-1}^{1}\frac{dz}{2}
\frac{\tilde{\Delta}_m}
{[(\tilde{\omega}_m-isz)^2+\tilde{\Delta}_m^2]^{1/2}} \nonumber \\
\quad +  \left(\tau_N^{-1}- \tau_P^{-1}\right)
\int_{-1}^{1}\frac{dz}{4}
\frac{\tilde{\Delta}_n}
{[(\tilde{\omega}_n-isz)^2+\tilde{\Delta}_n^2]^{1/2}},
\label{gap}
\end{gather}
\begin{gather}
\tilde{\omega}_n= \omega_n
+ \frac{\pi T}{2} \sum_m
\lambda(n-m)\int_{-1}^{1}
\frac{(\tilde{\omega}_m-isz)dz}
{[(\tilde{\omega}_m-isz)^2+\tilde{\Delta}_m^2]^{1/2}}
\nonumber \\
\quad + \left(\tau_N^{-1}+\tau_P^{-1}\right)
\int_{-1}^{1}\frac{dz}{4}
\frac{\tilde{\omega}_n-isz}
{[(\tilde{\omega}_n-isz)^2+\tilde{\Delta}_n^2]^{1/2}}.
\label{omega}
\end{gather}
Here $\omega_n = \pi T (2n + 1)$ are the Matsubara frequencies, $n$ are integers, $\tau_N$ and $\tau_P$ are the scattering times on nonmagnetic and magnetic impurities, respectively,  $T$ is the temperature in energy units ($k_B=\hbar=1$), 
and $\lambda(n-m)$ describes electron-phonon interaction: 
\begin{gather}
\lambda(n - m) = \int_0^\infty 
\frac{2\nu\,\alpha^2F(\nu)d\nu}
{\nu^2 +(\omega_m - \omega_n)^2},
\label{eq:kernel}
\end{gather}
where $\alpha^2F(\nu)$ is the electron-phonon spectral function. The Coulomb pseudopotential $\mu^\ast$ is included with a cutoff frequency $\omega_c$ in the step function $\theta(x)=1$ at $x>0$ and $\theta(x)=0$ at $x<0$.
The effect of superflow is accounted for by the Doppler shift parameter
\begin{equation}
s = p_F v_s, 
\label{eq:superflow}
\end{equation}
where $p_F=\hbar k_F$ is the Fermi momentum, $v_s=(\hbar/2m^*)(\partial_x\varphi+2\pi A_x/\phi_0)$ is the superfluid velocity, $\varphi$ is the phase of the order parameter, $A_x$ is the vector-potential, $\phi_0$ is the magnetic flux quantum and $m^*$ is the effective electron mass. The integration over $z$ comes from the averaging of components of electron momenta parallel to current. It is convenient to introduce the dimensionless scattering parameters:
\begin{equation}
p = (2\pi T_c\tau_N)^{-1},
\qquad
p_1 = (2\pi T_c\tau_P)^{-1}.
\end{equation}
The cases $p\ll 1$ and $p>1$ correspond to the clean and the dirty limits, respectively. In this paper we disregard magnetic scattering and set $p_1=0$. 

The current density is given by \cite{NC}:
\begin{equation}
J=-\frac{3eN}{2k_F}\pi T\sum_m\mbox{Im}\!\int_{-1}^1\!\frac{(\tilde{\omega}_m-izs)zdz}{[(\tilde{\omega}_m-isz)^2+\tilde{\Delta}_m^2]^{1/2}},
\label{J}
\end{equation}
where $N$ is the electron density and $e$ is the electron charge. At weak currents $s\ll \tilde{\Delta}$, the linearized Eq. (\ref{J}) yields $J=-A/\mu_0\lambda_L^2$ defining the magnetic penetration depth $\lambda_L$ in the local limit $\kappa\gg 1$:
\begin{equation}
\frac{\lambda_{L0}^2}{\lambda_L^2}=2\pi T\sum_{n=0}^\infty\frac{\tilde{\Delta}_n^2}{(\tilde{\Delta}_n^2+\tilde{\omega}_n^2)^{3/2}},
\label{lond}
\end{equation}
where $\lambda_{L0}=(m^*/\mu_0e^2N)^{1/2}$ is the London penetration depth in a clean superconductor at $T=0$.

\subsection{Calculation of $B_s$ and $B_c$ from $J(Q)$.}

At $\kappa\gg 1$ the superheating and thermodynamic critical fields can be obtained from the local relation $J(Q)$ as follows. Let a uniform magnetic induction  $B_0$ be applied along the $y$ axis parallel to a planar surface at $x=0$ and the screening gauge-invariant vector potential $Q(x)=A+\phi_0\varphi'/2\pi$ decreases slowly over the coherence length $\xi$. Then the Maxwell equations read: 
\begin{equation}
\frac{d^2 Q}{dx^2}=\mu_0J(Q).
\label{eq:dHdx}
\end{equation}
Following Ref. \cite{LG}, we calculate $B_s$ by multiplying Eq.~\eqref{eq:dHdx} by $dQ/dx$ and integrating from $x=0$ to $x=\infty$ using the boundary conditions $dQ/dx=0$ at $x=\infty$ and $dQ/dx=B_0$ and $Q(x)=Q_0$ at $x=0$. This yields:
\begin{equation}
B_0^2 = 2\mu_0\!\int_0^{Q_0}\! J(Q)dQ.
\label{B0}
\end{equation}
The superheating field $B_s$ is obtained from the condition that $Q_0$ at the surface reaches the depairing limit corresponding to the maximum $J(Q_c)$ at $Q_0=Q_c$:
\begin{equation}
B_s^2 = 2\mu_0\!\int_0^{Q_c}\!J(Q)dQ.
\label{Bs}
\end{equation}
This equation can be recast into a more transparent form using that $J=\partial {\cal F}/\partial Q$, where ${\cal F}$ is the free energy functional minimized by solutions of the Eliashberg equations. Furthermore,
\begin{equation}
J(Q)=\frac{\partial \cal{F}}{\partial Q}=\frac{d\cal{F}}{dQ}
\label{der}
\end{equation}
because $\tilde{\Delta}_n$ and $\tilde{\omega}_n$ satisfy Eqs. (\ref{gap}) and (\ref{omega}) which follow from the variational conditions $\delta {\cal F}/\delta \tilde{\Delta}_n=0$ and $\delta {\cal F}/\delta\tilde{\omega}_n=0$ ~\cite{C}. As a result, Eq. (\ref{Bs}) can be integrated to give:
\begin{equation}
B_s^2=2\mu_0(F_0-F_c).
\label{energ}
\end{equation}   
Here the condensation free energy $-F_0$ of superconducting state with no current $(Q=0)$ and the condensation energy $-F_c$ in a current-carrying  state with the critical velocity $v_c=\pi\hbar Q_c/m^*\phi_0$ are obtained from $\cal{F}$ expressed in terms of solutions of Eqs. (\ref{gap}) and (\ref{omega}).  Setting $Q_0$ in Eq. (\ref{B0}) to $Q_m$ at which the superfluid density, $J(Q_m)$ and  $F(Q_m)$ vanish, reduces Eq. (\ref{B0}) to the definition of the thermodynamic critical field:
\begin{equation}
B_c^2 = 2\mu_0\!\int_0^{Q_m}\!J(Q)dQ = 2\mu_0F_0.
\label{Bc}
\end{equation}
Here $F_0$ is given by \cite{C}:
\begin{gather}
F_0 = 2\pi N(0) T \sum_{n=0}^\infty
\left[ \sqrt{\omega_n^2 + \Delta^2_n} - \omega_n \right]\times
\nonumber\\
\left[ Z_S - 
\frac{Z_N\omega_n}{\sqrt{\omega_n^2 + \Delta^2_n}}\right],
\label{F0}
\end{gather}
where $Z_{N,S}=\tilde{\omega}_n/\omega_n$ are renormalization factors in the superconducting and normal states, respectively, and $N(0)$ is the density of states at the Fermi surface. The energy of the current-carrying state $F_c$ is:
\begin{gather}
\!\!F_c=\pi N(0) T\mbox{Re}\!\! \int_{-1}^1\!\!\frac{dz}{2}\sum_n
\bigg[ \sqrt{(\tilde{\omega}_n-isz)^2 + \tilde{\Delta}^2_n} - 
\nonumber \\
\!\!(\tilde{\omega}_n-isz)\mbox{sign}(\omega_n)\bigg]\!
\bigg[ Z_S - 
\frac{Z_N(\tilde{\omega}_n-isz)}{\sqrt{(\tilde{\omega}_n-isz)^2 + \tilde{\Delta}^2_n}}\bigg].
\end{gather}

\begin{figure}[!t]
\centering
\includegraphics[width=0.50\textwidth]{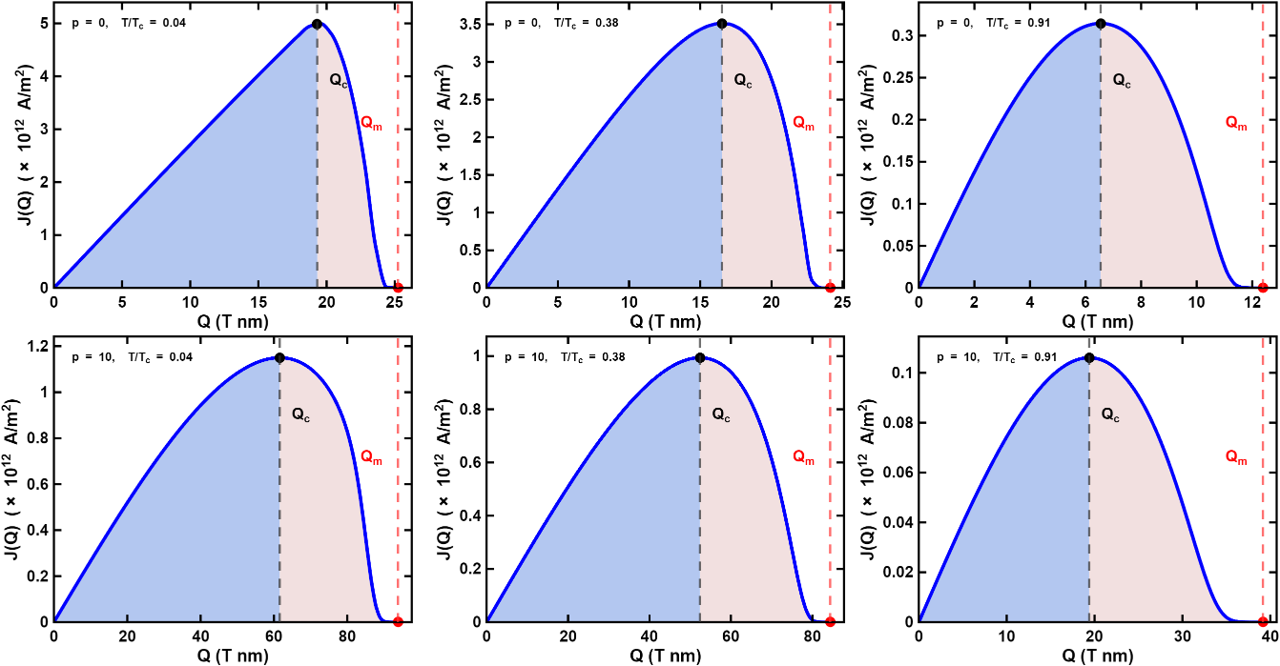}
\caption{Supercurrent density $J(Q)$ at different $T/T_c$ calculated for Nb$_3$Sn in the clean $(p=0)$ and the dirty $(p=10)$ limits represented by the top and the bottom panels. The blue and pink regions define the stable ($0 \leq Q \leq Q_c$) and unstable ($Q_c \leq Q \leq Q_m$) branches of $J(Q)$. Dashed lines mark the critical momentum $Q_c$ and the terminal momentum $Q_m$. Here $B_s^2$ and $B_c^2$ are determined by the area of the blue region and the total area under the $J(Q)$ curve, respectively.}
\label{F1}
\end{figure}

Using Eqs. (\ref{Bs}) and (\ref{Bc}), we obtain:
\begin{equation}
\frac{B_s}{B_c}=
\left[
\frac{\int_0^{Q_c} J(Q)dQ}
{\int_0^{Q_m} J(Q)dQ}
\right]^{1/2}\!\!=\left(1-\frac{F_c}{F_0}\right)^{1/2}.
\label{ratio}
\end{equation}
Shown in Fig. \ref{F1} are examples of $J(Q)$ calculated for clean and dirty Nb$_3$Sn described in detailed in the next section. Here $(B_s/B_c)^2$ equals the ratio of the respective areas under the $J(Q)$ curves, where $Q_c$ determines the depairing current density and $Q_m$ defines the terminal superfluid velocity at which superconductivity is suppressed. At $\kappa\gg 1$ the superheating field given by Eqs. (\ref{ratio}) is always smaller than the thermodynamic critical field. Because $B_c$ of a superconductor with isotropic Fermi surface is independent of scattering on nonmagnetic impurities, the integral in Eq. (\ref{Bc}) is independent of $p$, even though both $J(Q)$ and $Q_m$ depend on $p$. As follows from Fig. \ref{F1}, $J(Q)$ in the clean limit at $T\ll T_c$ remains nearly linear up to very close to $Q_c$. This shows that the lack of the nonlinear Meissner effect in the BCS theory in the clean limit at $T=0$ \cite{nlme} remains valid in the Eliashberg theory as well. In this case the shaded area in the top left Fig. \ref{F1} can be approximated by a triangle, giving $B_s\approx(\mu_0J_cQ_c)^{1/2}$ and:
\begin{equation}
B_s/B_c\approx (Q_c/Q_m)^{1/2},\quad p=0,\quad T=0.
\label{cl}
\end{equation} 

Taking $J\propto Q[1-(Q/Q_m)^2]$ and $Q_c=Q_m/\sqrt{3}$ at $T\approx T_c$ in Eq. (\ref{ratio}) reproduces the GL result~\cite{sh1,Galaiko}:
\begin{equation}
B_s/B_c
=\sqrt{5}/3\approx 0.745,\quad T\approx T_c
\label{GL}
\end{equation}
Equations (\ref{Bs}) and (\ref{Bc}) can be used to calculate $B_s(T)/B_c(T)$ numerically in the BCS model for different scattering parameters $p$ using the free energy functional of the Eilenberger theory ~\cite{eilenb,LG}. In the clean BCS limit $B_s(0)$ at $T=0$ and $\kappa\gg 1$ was obtained by Galaiko  \cite{Galaiko}: 
\begin{equation}
B_s(0)/B_c(0) \approx 0.84,  \qquad  p=0
\label{BCS_clean}
\end{equation}
In the BCS dirty limit with $p= 10-20$ the ratio $B_s(0,p)/B_c$ is close to $0.8$ \cite{LG} and approaches \cite{kubo}:
\begin{equation}
B_s(0)/B_c(0) \to 0.795,  \qquad  p\to\infty
\label{BCS_dirty}
\end{equation}
In what follows we use Eq. (\ref{ratio}) to calculate $B_s(T)$ from the Eliashberg theory to see how the effects of strong electron-phonon coupling, low-frequency phonons and nonmagnetic impurity scattering changes $B_s/B_c$ as compared to the BCS results.

\vspace{-1mm}
\section{Numerical results}
\begin{figure}[H]
\centering
\includegraphics[width=0.450\textwidth]{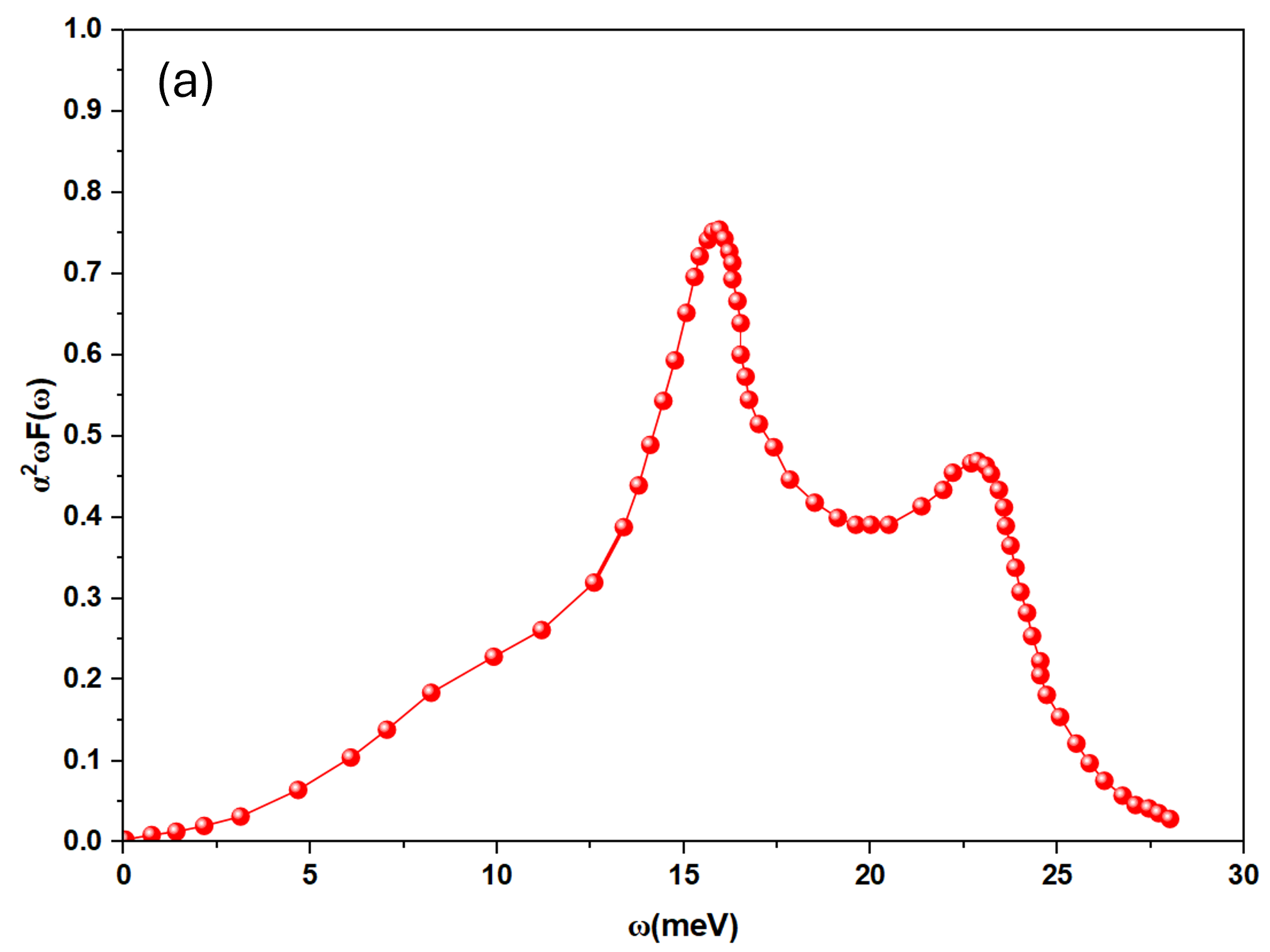}
\includegraphics[width=0.450\textwidth]{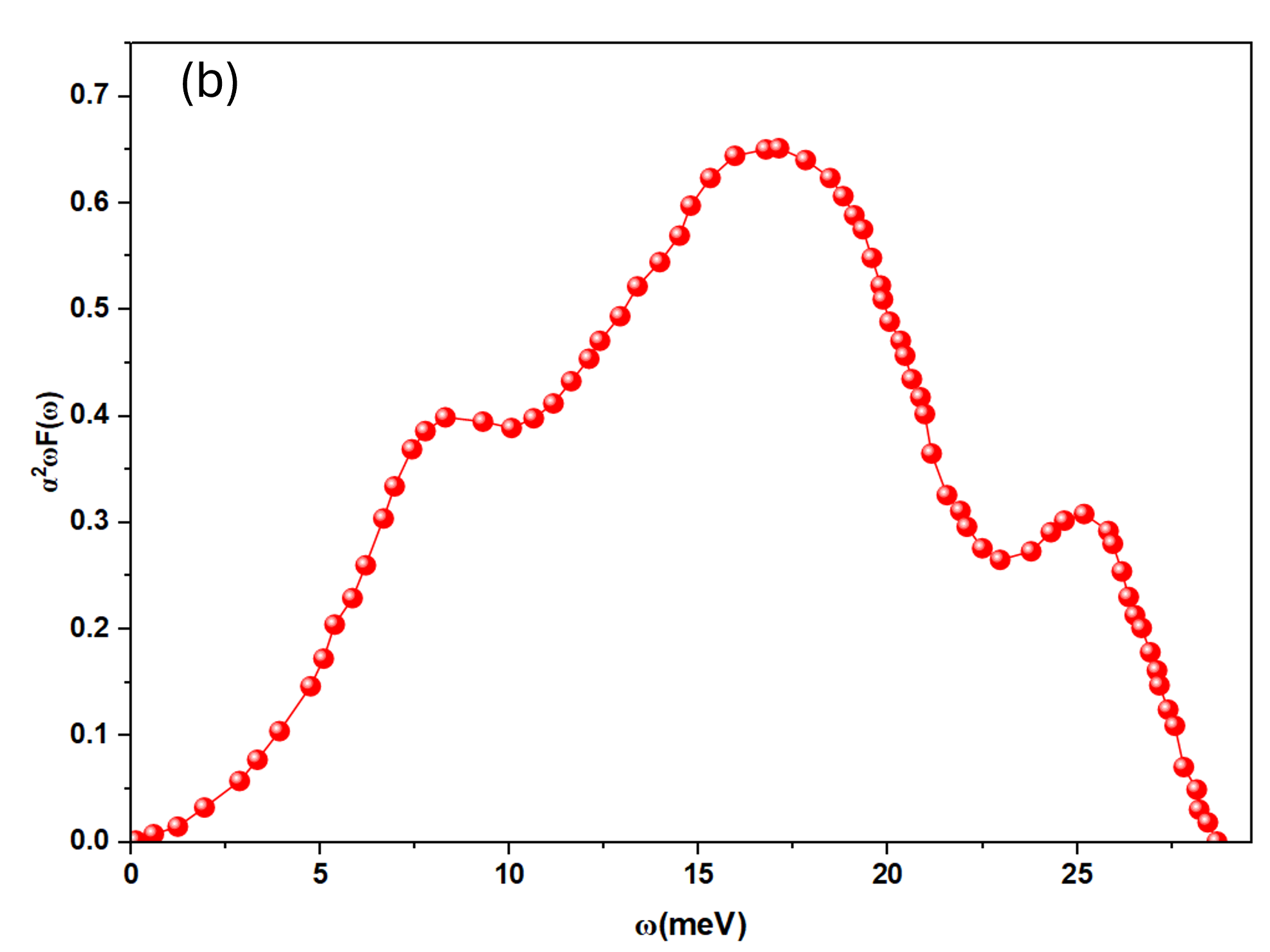}
\caption{The electron-phonon spectral functions $\alpha^2F(\omega)$ for: (a) Nb \cite{Arnold2} and (b)  Nb$_3$Sn \cite{Shen1972} used for the calculations of $B_s(T,p)$ and $B_c(T)$.}
\label{F2}
\end{figure}

In this section we present numerical calculations of $B_s(T,p)$ as functions of temperature and the nonmagnetic scattering rate for Nb and Nb$_3$Sn using their electron-phonon spectral functions $\alpha^2F(\omega)$ extracted from tunneling experiments~\cite{Arnold1, Arnold2, Shen1972}. For these $\alpha^2F(\omega)$ shown in Fig. \ref{F2}, we computed the electron-phonon coupling constants $\lambda = 1.038$ for Nb and $\lambda=1.57$ for Nb$_3$Sn. Other relevant material parameters are listed in Table~I. 
\begin{table}[h]
\caption{Material parameters of Nb and Nb$_3$Sn.}
\begin{ruledtabular}
\begin{tabular}{lcc}
Parameter & Nb & Nb$_3$Sn \\
\hline
$T_c$ (K) & 9.20 & 18.20 \\
$\lambda$ & $1.038$~\cite{Arnold2} & $1.57$~\cite{Shen1972} \\
$\mu^\ast$ & 0.215 & 0.105 \\
$v_F$ (m/s) & $6.3\times10^5$ & $2.12\times10^5$ \\
$N$ (m$^{-3}$) & $2.8\times10^{29}$ & $3.0\times10^{29}$ \\
$k_F$ (m$^{-1}$) & $5.4\times10^9$ & $1.8\times10^{10}$ \\
$N(0)$ (J$^{-1}$m$^{-3}$) & $3.40\times10^{47}$~\cite{Beck1970} & $6.8\times10^{47}$~\cite{Mitrovic1984} \\
\end{tabular}
\end{ruledtabular}
\label{tab:parameters}
\end{table}

\subsection{Nb}

\begin{figure}[H]
\centering
\includegraphics[width=0.450\textwidth]{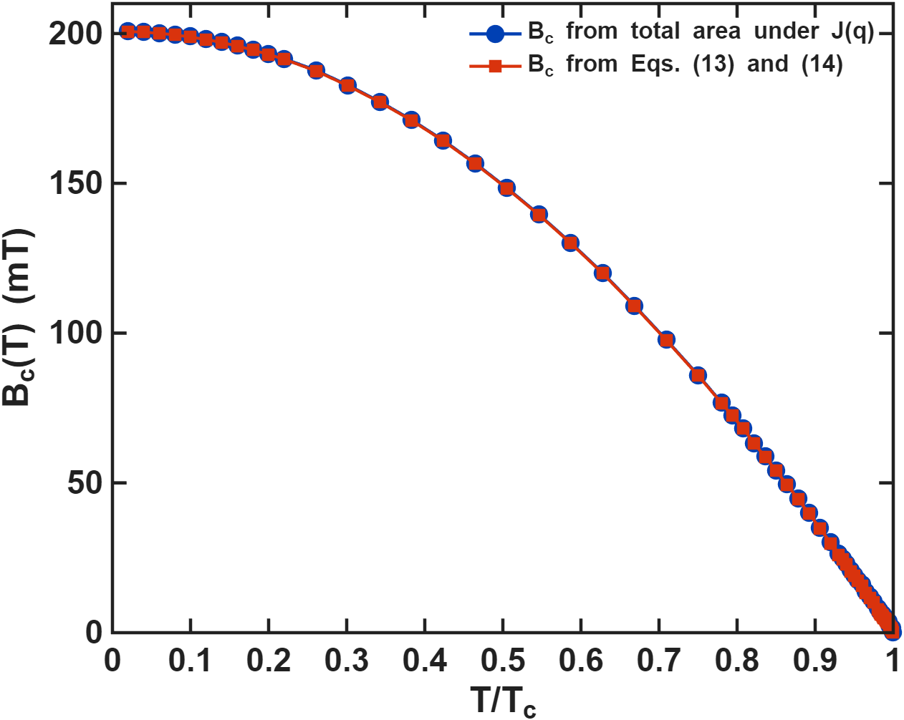}
\caption{Thermodynamic critical field for Nb, where the solid line is calculated from Eqs. (\ref{Bc}) and (\ref{F0}) and the dots are obtained by integrating $J(Q)$, as described in the text.}
\label{F3}
\end{figure}

\begin{figure}[H]
\centering
\includegraphics[width=0.480\textwidth]{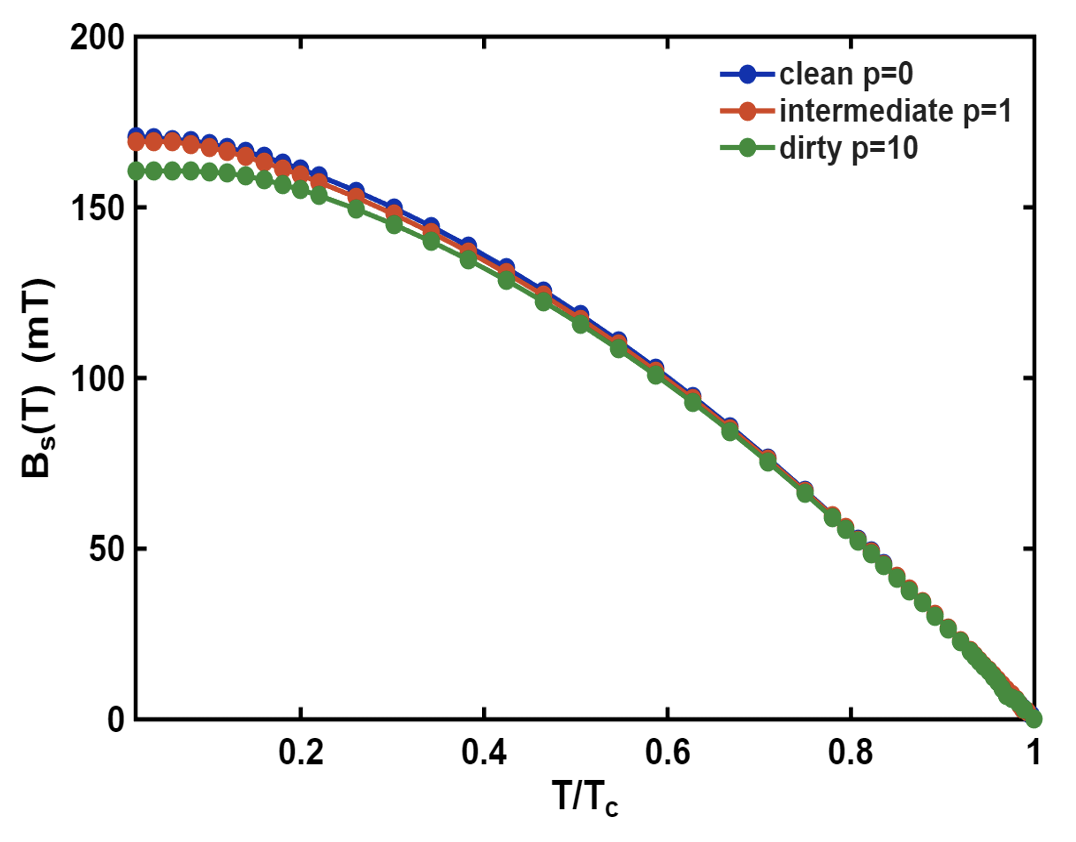}
\caption{Superheating field for Nb for different scattering parameters calculated by integrating $J(Q)$ in Eq. (\ref{Bs}).}
\label{F4}
\end{figure}

We solved the Eliashberg equations numerically as described in Appendix A and calculated $B_s(T,p)$ and $B_c(T)$ from Eqs. (\ref{Bs}), (\ref{Bc}) and (\ref{ratio}) for the above parameters of Nb. The results are shown in Figs. \ref{F3} and \ref{F4}. Here $B_c$ calculated from the area under the $J(Q)$ curve matches $B_c$ obtained from Eqs. (\ref{Bc}) and (\ref{F0}) and reproduces the conventional value $B_c(0)= 200$ mT for Nb \cite{C}. Shown in Fig. \ref{F4} is $B_s(T,p)$ calculated for different dimensionless scattering rates $p$. The so-obtained $B_s$ does not represent the actual superheating field of a clean Nb which is a marginal type-II superconductor with $\kappa\approx 1$ for which calculations of $B_s$ requires a stability analysis of the Meissner state with respect to inhomogeneous perturbations $\delta\Delta_n({\bf r})$ and $\delta {\bf Q}({\bf r})$ \cite{sh1,sh2,sh3,sh4}, which is beyond the scope of this work. The cases of $p=0$ and $p=1$ are included here to show the generic evolution of $B_s(T,p)$ as $p$ is increased. Yet at $p=10$, the GL parameter $\kappa\gtrsim 10$ becomes large enough for Eq. (\ref{Bs}) to be applicable.

\begin{figure}[H]
\centering
\includegraphics[width=0.50\textwidth]{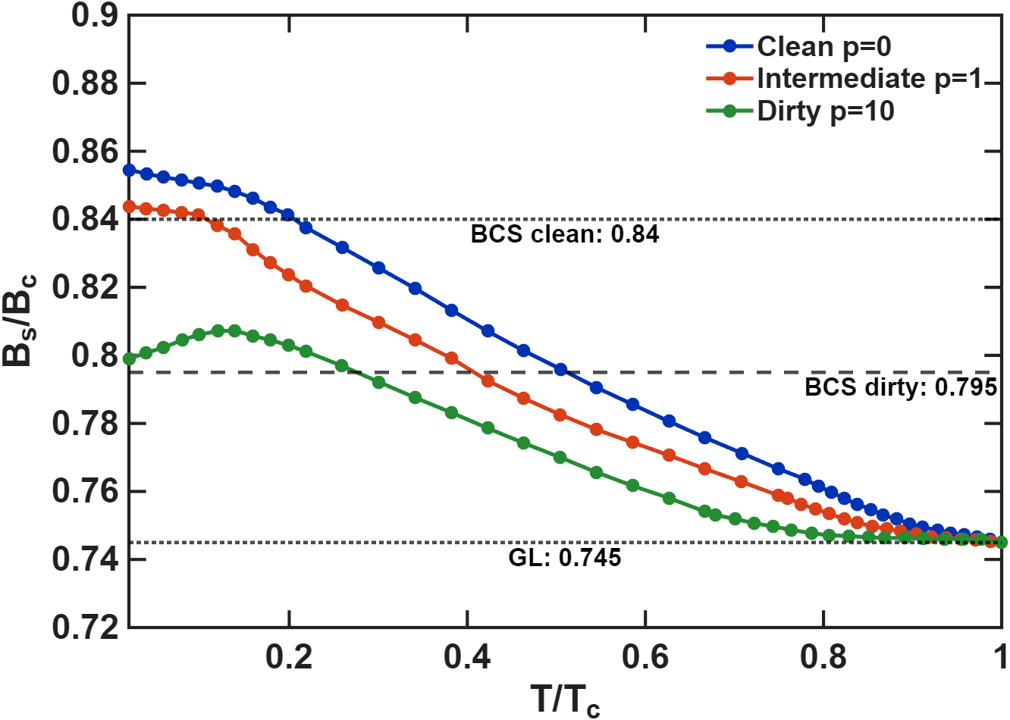}
\caption{The ratio $B_s/B_c$ for Nb at different impurity scattering parameters. Dashed lines show the BCS results for the clean \cite{Galaiko} and the dirty \cite{kubo} limits at $T=0$ along with the GL result at $T\approx T_c$. }
\label{F5}
\end{figure}

\begin{figure}[H]
\centering
\includegraphics[width=0.50\textwidth]{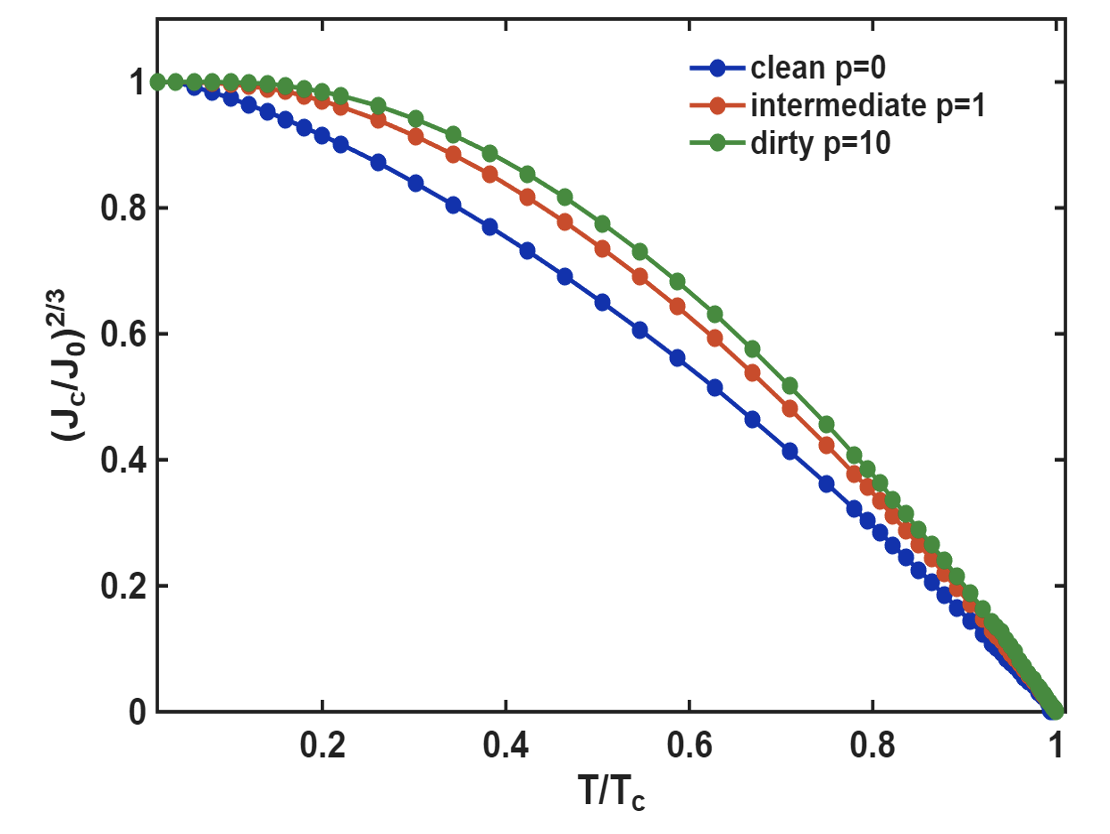}
\caption{The plot of $[J_c/J_c(0)]^{2/3}$ versus $T/T_c$ for different scattering parameters $p$. The linear behavior near $T_c$ is consistent with the GL scaling, while deviation from the linear behavior in the clean limit at lower temperatures ~\cite{KL} is indicative of the peaibreaking effect of low-energy phonons.}
\label{F6}
\end{figure}

The ratio $B_s/B_c$ shown in Fig. \ref{F5} exhibits the features characteristic of a type-II superconductor which are discussed below for Nb$_3$Sn in more detail. Here we only emphasize a few essential points: 1. $B_s/B_c$ calculated from the Eliashberg theory for the realistic parameters of Nb {\it exceeds} the BCS prediction in the clean limit. 2. $B_s/B_c$ decreases as the scattering parameter $p$ is increased beyond 1,  similar to the results of the Eilenberger theory  \cite{LG}. 3. In the dirty limit $(p =10)$, the ratio $B_s/B_c\approx 0.8$ turns out to be close to the BCS results \cite{LG,kubo}.   4. A maximum in $B_s(T)/B_c(T)$ develops upon increasing $p$.  
Having calculated multiple $J(Q)$ curves at different $T$ and $p$ like those shown in Fig. \ref{F1}, we obtained the depairing current density $J_c(T,p)$  shown in Fig. \ref{F6}.  This behavior of $J_c(T,p)$ is consistent with the results of Eilenberger theory \cite{KL} and Eliashberg theory for a model spectral function $\alpha^2F(\omega)$ \cite{NC}. Particularly, we observed the GL dependence $J_c\propto (1-T/T_c)^{3/2}$ near $T_c$ and $J_c(T,B)$ decreasing with $p$.  The temperature dependencies of $J_c(T)$ at $T\ll T_c$ in the clean and the dirty limits turned out to be rather different. In the dirty limit the nearly flat $J_c(T)$ shown in Fig. \ref{F6} is similar to that of the BCS theory~ \cite{KL}, while $J_c(T)$ in the clean limit can be described by a power-law dependence $J_c(0)-J_c(T)\propto T^n$ with $n\approx 1.5$. This difference reflects a peculiar effect of current on the quasiparticle density of states in the presence of nonmagnetic impurities~\cite{Fulde,LG}. For instance, a clean BCS superconductor at $J=J_c$ is in a gapless state in which $J_c(T)=J_c(0)(1-T/T_c)^{3/2}$ at all $T$ giving a linear dependence of $J_c(0)-J_c(T)\propto T$ at $T\ll T_c$  \cite{KL}.  As $p$ is increased, a quasiparticle energy gap $E_g$ at $J=J_c$ opens at $p>p_c\simeq 0.1$ and approaches $E_g\approx 0.32\Delta_0$ at $p\gg 1$ \cite{LG}. As a result, the temperature dependence of $J_c(T)$ at $T\ll T_c$ crosses over from the power-law in the gapless clean limit $p\lesssim p_c$ to the thermally-activated behavior at $p>1$ due to gapped quasiparticles in the dirty limit. Our results suggest that the behavior of $J_c(T)$ is affected by subgap quasiparticles caused by low-frequency phonons which presumably result in the deviation from the linear temperature dependence of $[J_c(T)/J_c(0)]^{2/3}$ of the Eilenberger theory in the clean limit \cite{KL}, as shown in Fig. \ref{F6}.

\begin{figure}[H]
\centering
\includegraphics[width=0.50\textwidth]{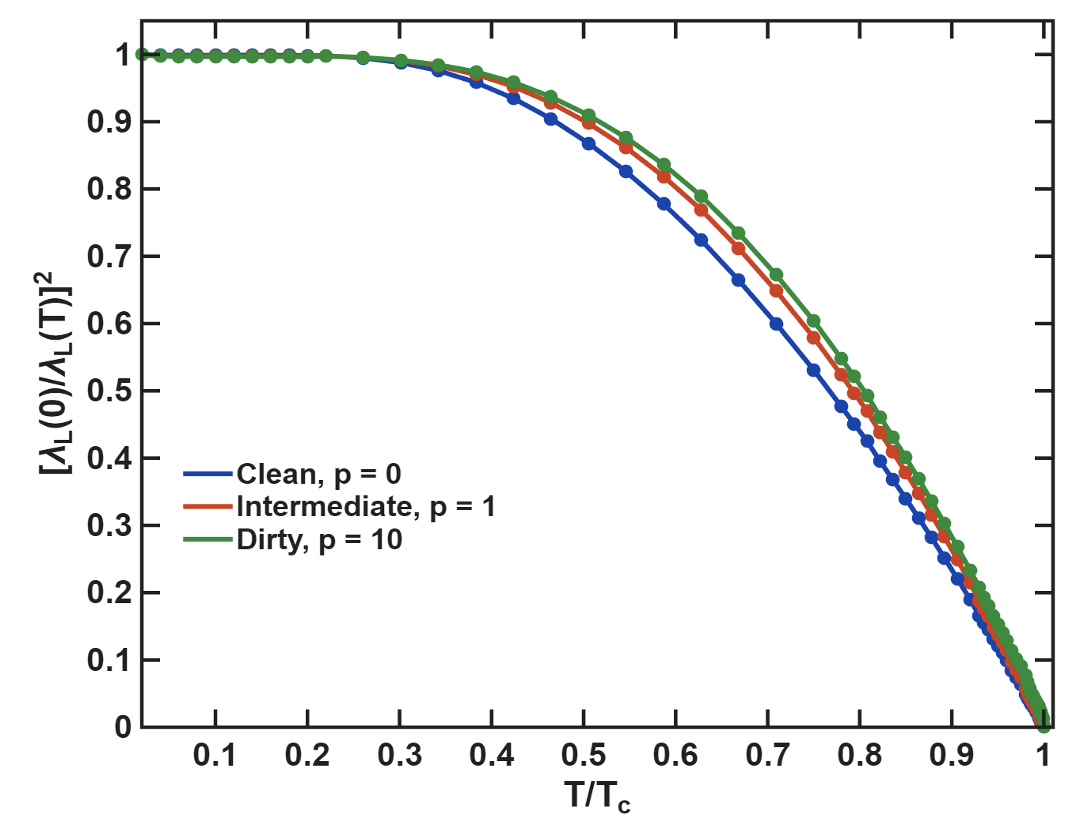}
\caption{The plot of $[\lambda_L(0)/\lambda_L(T)]^2$ versus $T/T_c$ for different scattering parameters $p$. The linear behavior near $T_c$ is consistent with the GL scaling.}
\label{F7}
\end{figure}

To complete the characterization of Nb, Fig. \ref{F7} shows the temperature dependencies of  $[\lambda_L(0)/\lambda_L(T,p)]^2$ quantifying the normalized superfluid densities at different scattering parameters.  All $[\lambda_L(0)/\lambda_L(T,p)]^2$ curves exhibit the linear GL behavior near $T_c$ and flatten at low $T$, consistent with the behavior of $\lambda_L(T)$ in s-wave superconductor in the Eliashberg theory \cite{C,MC}.  

\subsection{Nb$_3$Sn}
\begin{figure}[H]
\centering
\includegraphics[width=0.50\textwidth]{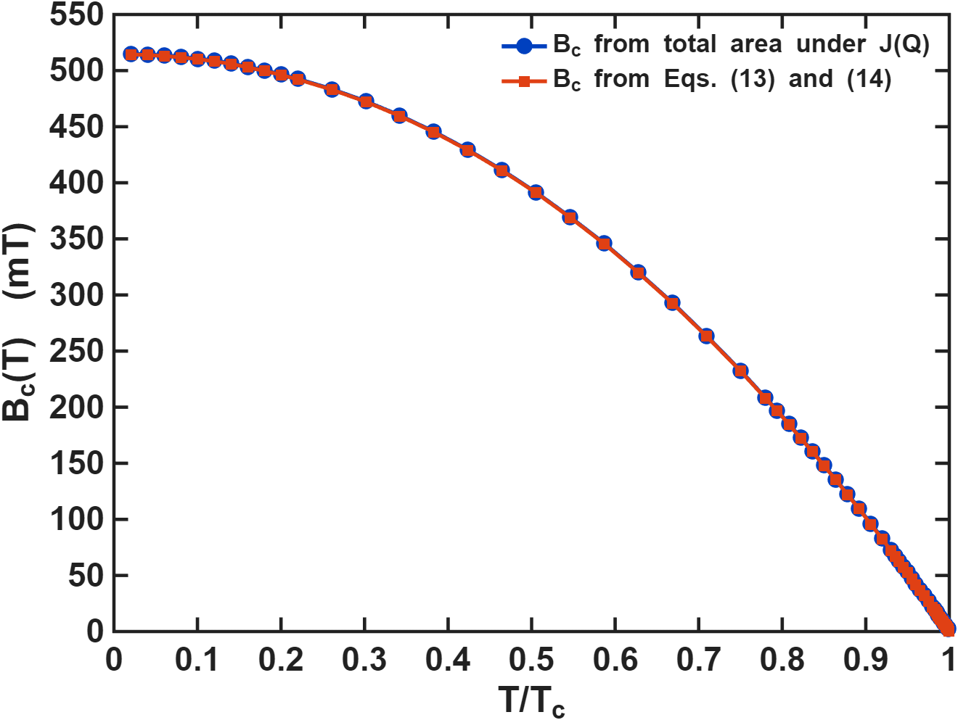}
\caption{Thermodynamic critical field for Nb$_3$Sn, where the solid line is calculated from Eqs. (\ref{Bc}) and (\ref{F0}) and the dots are obtained by integrating $J(Q)$, as described in the text.}
\label{F8}
\end{figure}

\begin{figure}[H]
\centering
\includegraphics[width=0.50\textwidth]{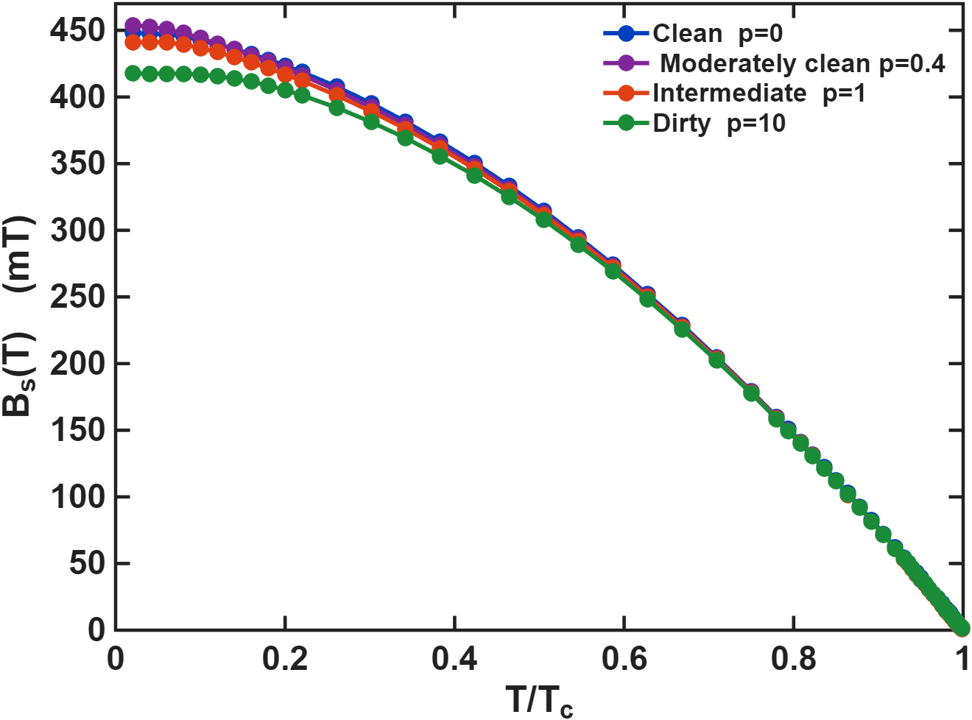}
\caption{Superheating field for Nb$_3$Sn for different scattering parameters calculated by integrating $J(Q)$ in Eq. (\ref{Bs}).}
\label{F9}
\end{figure}

 $B_s(T,p)$ and $B_c(T)$ of Nb$_3$Sn were calculated for the parameters listed in Table 1 and $\alpha^2F(\omega)$ shown in Fig. \ref{F2}(b). The results are shown in Figs. \ref{F8} and \ref{F9}. Here $B_c$ calculated from the area under the $J(Q)$ curve matches $B_c$ obtained from Eqs. (\ref{Bc}) and (\ref{F0}) and yields $B_c(0)\approx 514$ mT characteristic of a stoichiometric Nb$_3$Sn \cite{Mitrovic1984}.

\begin{figure}[H]
\centering
\includegraphics[width=0.50\textwidth]{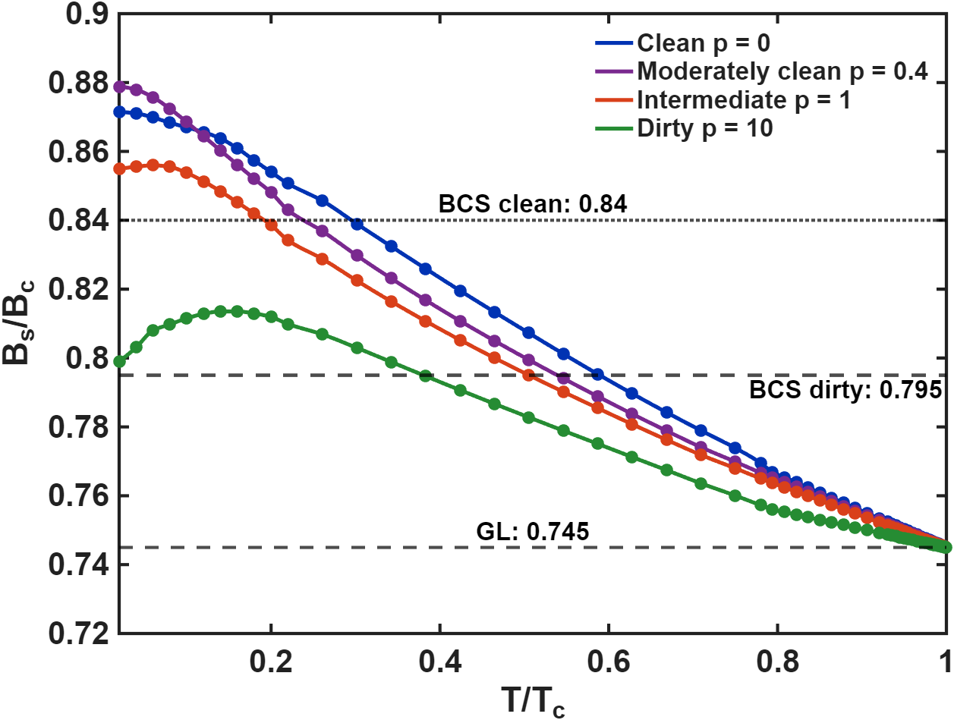}
\caption{The ratio $B_s/B_c$ for Nb$_3$Sn at different impurity scattering parameter $p$. Dashed lines show the BCS results for the clean \cite{Galaiko} and the dirty \cite{kubo} limits at $T=0$ along with the GL result at $T\approx T_c$.}
\label{F10}
\end{figure}

The ratio $B_s/B_c$ shown in Fig. \ref{F10} exhibits the same generic features discussed above for Nb, however $B_s/B_c$ calculated for Nb$_3$Sn {\it exceeds} the BCS results to a greater extent than for Nb, consistent with the larger electron-phonon coupling constant for Nb$_3$Sn. Because even a clean Nb$_3$Sn is a true type-II superconductor, we present here detailed calculations of $B_s(T,p)$ as a function of the scattering parameter $p$, from the clean to the dirty limit.  Given strong dependencies of superconducting properties of Nb$_3$Sn on Sn content, we apply the Eliashberg theory to a stoichiometric Nb$_3$Sn with maximum $T_c$ assuming that it contains passive atomic impurities other than Sn or Nb which cause scattering but do not affect $T_c$ \cite{orlando,godeke}.  

\begin{figure}[H]
	\centering
	\includegraphics[width=0.50\textwidth]{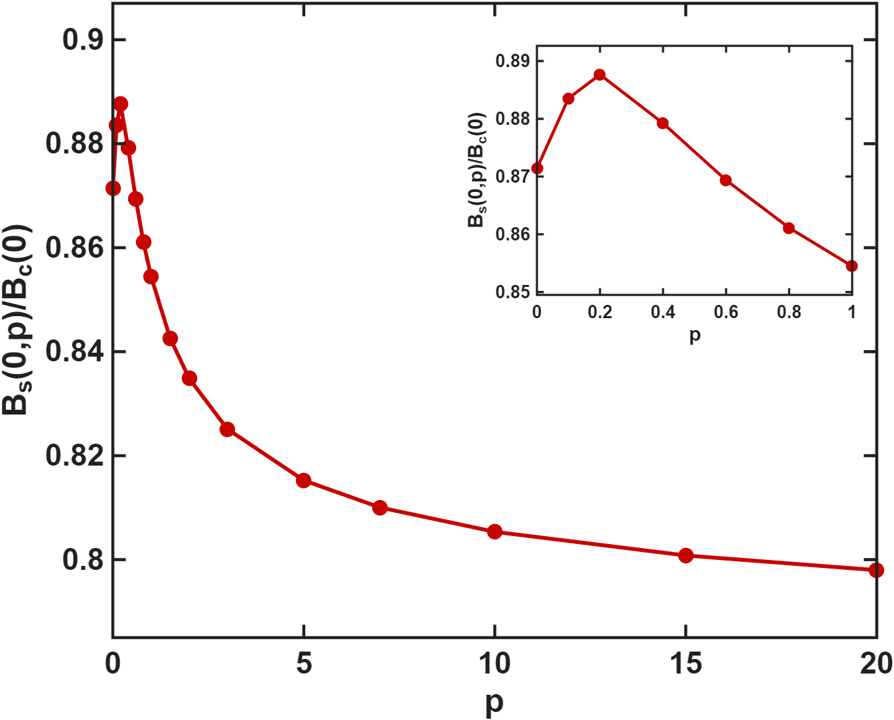}
	\caption{The ratio $B_s(0,p)/B_c(0)$ for Nb$_3$Sn as a function of the impurity scattering parameter $p$ at $T=0$. Insert shows the behavior of $B_s(0,p)/B_c(0)$ at the maximum. }
	\label{F11}
\end{figure}

Figure \ref{F10} shows intricate effects of impurity scattering on $B_s(T)$ which manifesting themselves in a maximum in $B_s(T)$ at low temperatures emerging upon increasing $p$ and crossing $B_s(T)$ curves with different $p$. The latter gives rise to a non-monotonic dependence of the ratio $B_s(0,p)/B_c(0)$ on the impurity scattering rate $p$ at $T=0$  with a maximum $B_s(0,p)/B_c(0)\approx 0.888$ at $p\approx 0.2$, as shown in Fig. \ref{F11}. This behavior of $B_s(T,p)/B_c(T)$ is qualitatively similar to that calculated from the Eilenberger theory \cite{LG}.   For instance, the maximum in $B_s(0,p)/B_c(0)$ shown in Fig. \ref{F11} appears in the range of $p$ where the quasiparticle gap at $B=B_s$ in the Eilenberger theory opens up upon increasing $p$ \cite{LG}. This maximum appears in the Eliashberg theory as well despite the broadening of the gap peaks in the density of states by inelastic scattering on phonons.  Yet we do not observe a small maximum in $B_s(T)$ in the clean limit obtained from the Eilenberger theory \cite{Catelani}. This maximum in $B_s(T)$ in the clean limit may be washed out by the subgap quasiparticle states caused by scattering on phonons but it emerges as $p$ is increased.

\begin{figure}[H]
\centering
\includegraphics[width=0.48\textwidth]{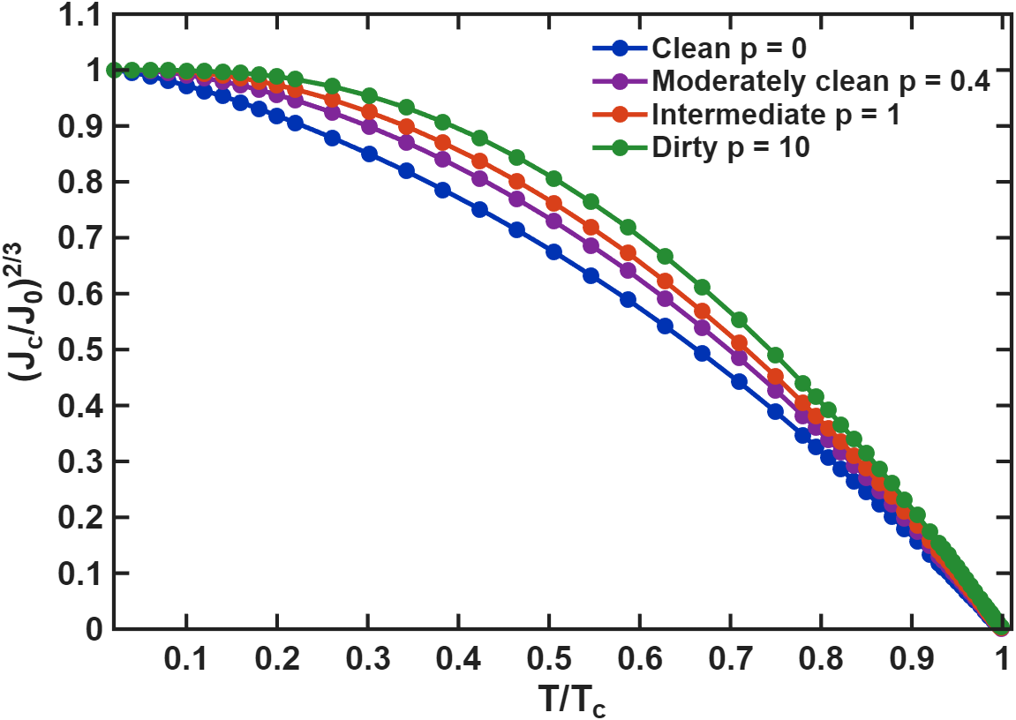}
\caption{The plot of $[J_c/J_c(0)]^{2/3}$ versus $T/T_c$ for different scattering parameters $p$. The linear behavior near $T_c$ is consistent with the GL scaling, while deviation from the linear behavior in the clean limit at lower temperatures ~\cite{KL} is indicative of the paibreaking effect of low-energy phonons.}
\label{F12}
\end{figure}

\begin{figure}[H]
\centering
\includegraphics[width=0.48\textwidth]{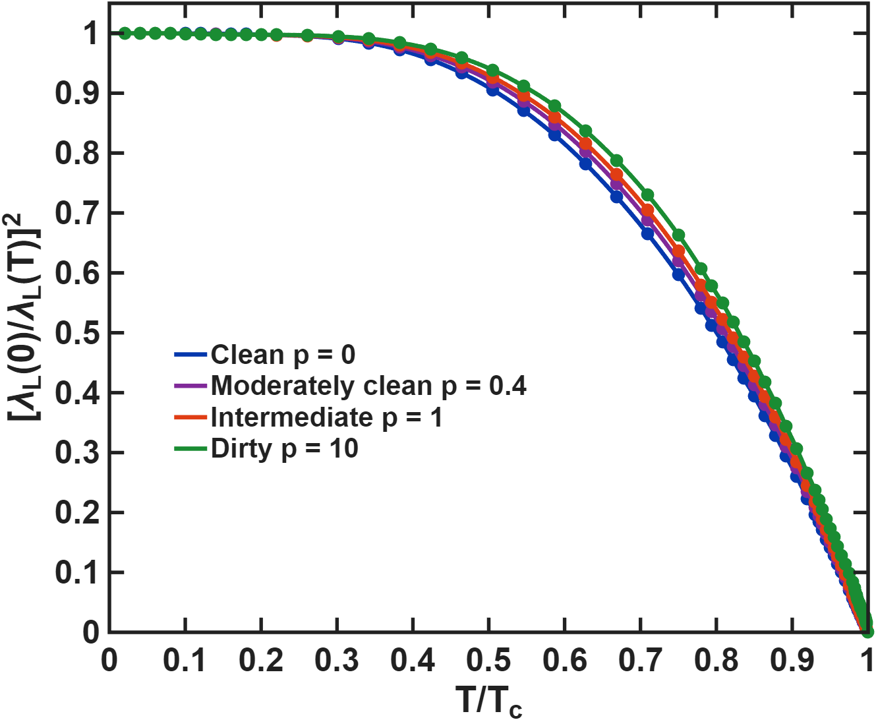}
\caption{The plot of $[\lambda_L(0)/\lambda_L(T)]^2$ versus $T/T_c$ for different scattering parameters $p$. The linear behavior near $T_c$ is consistent with the GL scaling.}
\label{F13}
\end{figure}

In conclusion of this section, we show in Fig. \ref{F12} and \ref{F13} temperature dependencies of $[J_c(T,p)/J_0(p)]^{3/2}$ and  $[\lambda_L(0)/\lambda_L(T,p)]^2$ calculated for Nb$_3$Sn at different scattering parameters. These results for Nb$_3$Sn are qualitatively similar to those shown in Figs. \ref{F6} and \ref{F7} for Nb.   

\section{Conclusions}

In this work we calculate the static superheating field from the Eliashberg theory for extreme type-II superconductors with $\kappa\gg 1$ and intermediate strength of electron phonon coupling and show that it consistently exceeds $H_s(T)$ predicted by the BCS theory. We show that interplay of strong coupling effects and the subgap states coming from finite quasiparticle lifetimes results in the net increase of $H_s$ as compared to the BCS results, the enhancement of $H_s$ is most pronounced in cleaner superconductors with low concentration of nonmagnetic impurities. Our calculations for Nb$_3$Sn with the electron-phonon spectral function $\alpha^2F(\omega)$ extracted from tunneling measurements give a nonmonotonic dependence of $H_s(T,p)$ on the impurity scattering rate with a maximum at $p\approx 0.2$ at which the ratio $H_s(0)/H_c$ is about 6\% higher than that in the BCS clean limit \cite{Galaiko}.

The intricate behavior of $H_s(T,p)$ controlled by the impurity scattering rate comes from the evolution of the density of states $N(E,p)$ at $H=H_s$ upon increasing disorder. In the BCS model the quasipaticle gap $E_g$ in a clean superconductor closes at a field $H_g$ smaller $H_s$ resulting in a gapless state at $H=H_s$. As the scattering rate $p$ increases, $H_g$ becomes larger than $H_s$ and the quasiparticle gap appears at $H=H_s$. This result of the Eilenberger theory \cite{Fulde,LG} is qualitatively consistent with our Eliashberg calculations, although instead of a well-defined BCS quasiparticle gap at $H_s<H_g$, there appears a pseudo gap caused by the broadening of the density of states by inelastic electron-phonon scattering.

Calculations of $H_s$ at intermediate GL parameters $\kappa$ which requires the stability analysis of superconducting state with respect to inhomogeneous perturbations of current and the order parameter remains an outstanding problem in the Eliashberg theory, as well as the incorporation of the DFT to account for the effects the Coulomb interaction in actual electron bands \cite{ab1,ab2}.    Another unresolved issue important for applications \cite{accel1,accel2,Xlasers} is the applicability of the quasi-static $H_s$ as the breakdown field limit of a superconductor exposed to strong low frequency electromagnetic fields with $\hbar\omega\ll k_BT_c$.  Calculation of a dynamic superheating field $H_d$ in nonequilibrium BCS theory near $T_c$ showed that $H_d$ can exceed $H_s$ by a factor $\sqrt{2}$ if the RF period is much smaller than the electron-phonon energy relaxation time ~\cite{SG}. However at low temperatures the quasi-static $H_s$ may be applicable if a superconductor is not too clean $(p\gtrsim 0. 2)$ and a quasiparticle pseudogap at $H=H_s$ ensures a small density of nonequilibrium quasiparticles.

\begin{acknowledgments}
This work was supported by DOE under grant DE-SC0025560.
\end{acknowledgments}

\appendix
\section{Computational details}
The Eliashberg equations Eqs.~(\ref{gap}) and (\ref{omega}) for $\tilde{\Delta}_n$ and $\tilde{\omega}_n$ were solved self-consistently on the imaginary-frequency axis for the material parameters in Table~\ref{tab:parameters}. The electron--phonon spectral function $\alpha^2F(\omega)$ was interpolated onto a uniform 1200-point frequency grid using piecewise cubic Hermite interpolation. The Coulomb pseudopotential $\mu^{*}$ was applied below a cutoff frequency $\omega_c=10\omega_{\max}$, where $\omega_{\mathrm{max}}$ is the maximum frequency at which $\alpha^2F(\omega)$ shown in Fig. 2 is defined. The cutoff $\omega_{\mathrm{cut}}=12\omega_{\max}$ was applied in the pairing kernel. The Doppler-shifted angular average over the spherical Fermi surface was evaluated numerically at each self-consistency
step. The number of Matsubara frequencies was chosen to maintain the same $\omega_{\mathrm{cut}}$ for each $T$. Iterations were continued until the maximum relative change in both $\Delta_n$ and $Z_n$ between successive steps fell below $10^{-7}$--$10^{-9}$, with tighter tolerance near $T_c$. A maximum of 2200 (2600) iterations was used for the clean (dirty) limits. Linear mixing factors of $0.08$--$0.20$ were used to ensure numerical stability, and the converged solution at the nearest temperature or condensate momentum was used as the starting point for subsequent steps.

Equations~(\ref{gap}) and~(\ref{omega}) were solved iteratively at each $T$ and $s$, using 100 to 240 grid in $0<Q<Q_m$. The current density $J(Q)$ was calculated from the converged $\widetilde{\omega}_n$ and $\widetilde{\Delta}_n$ using Eq.~(\ref{J}), including both positive and negative-Matsubara frequencies. The resulting $J(Q)$ was calculated as a function of the gauge invariant vector potential $Q=A+\phi_0\varphi'/2\pi$; the critical current density $J_c(T)$ at $Q=Q_c(T)$ was determined from the maximum of $J(Q)$. Numerical integration over the nonuniform momentum grid used a generalized composite Simpson method. The superheating field $B_s(T)$ and thermodynamic critical field $B_c(T)$ were obtained by numerically integrating the respective areas under the $J(Q)$ curve, where $Q_c$ is the momentum at which the current density is maximal and $Q_m$ is the first zero of $J(Q)$ on the descending branch. As a consistency check, $B_c(T)$ was also calculated from Eqs. (\ref{Bc}) and (\ref{F0}), using superconducting and normal-state solutions evaluated on the same Matsubara-frequency grid. The London penetration depth was calculated from Eq.~(\ref{lond}) using the zero-current Eliashberg solutions over the positive $\omega_n$.

\end{document}